# Heterogeneous Turbulent Jets of the Mutually Immiscible Liquids: the Mixing and Heat Transfer Processes

Ivan Vasilievich Kazachkov[1,2]

[1]Department of Information Technology and Data Analysis, Nizhyn Gogol State University, 16600, UKRAINE,
[2]Department of Energy Technology, Royal Institute of Technology (KTH), 10044 Stockholm, SWEDEN
Ivan.Kazachkov@energy.kth.se

**Abstract –** Many natural and technical processes deal with the turbulent mixing and heat transfer in the jet of mutually immiscible liquids, which represent an important class of the modern multiphase systems dynamics. The differential equations for axisymmetrical two-dimensional stationary flow and the integral correlations in a cylindrical coordinate system are considered for the free and confined jets. The parameters of the turbulent mixing in a free jet and in two-phase flow in a chamber are modelled and analyzed. One example is also done for high-temperature flow of the liquid metal melt cooled by water in a simulation of the hypothetic severe accidents at the nuclear power plant is presented and compared to the experimental data. The algorithm and the results obtained may be of interest for some research and industrial tasks, where calculation of the parameters of multiphase turbulent mixing and heat transfer are important.

## 1. History of the problem and statement of the research

All characteristics $a^l(t)$ of the multiphase mixture (mass, velocity, temperature, etc.) in a multiphase flow are considered in accordance with the method of turbulent multiphase flows [1] developed and reported in many publications, e.g. [2-7]. With this approach, the analog of the Navier-Stokes equations in a boundary layer approximation was derived [1]:

$$\sum_{i=1}^{m}\left[\frac{\partial}{\partial x}(y\rho_i B_i u_i) + \frac{\partial}{\partial y}(y\rho_i B_i v_i)\right] = 0, \qquad \sum_{i=1}^{m} B_i = 1, \qquad (1)$$

$$\sum_{i=1}^{m} \rho_i B_i (u_i \frac{\partial u_i}{\partial x} + v_i \frac{\partial u_i}{\partial y}) = -\frac{dp}{dx} + \frac{1}{y}\frac{\partial}{\partial y}\left[y \sum_{i=1}^{m} B_i \tau_i\right]_m,$$

where are $a^l(t) = \sum_{i=1}^{m} B_i(t) a_i^l(t)$, $B_i(t)$ was introduced as co-called function-indicator of the phase (equal to 1 if a phase $i$ is present at the point and 0 – if a phase is absent at the considered point of the domain of a flow). In the stationary equations (1) for the flow of incompressible liquids, written in a cylindrical coordinate system, are: *p*- pressure, ρ- density, *u,v*- the longitudinal and transversal velocity components, $\tau_i$- turbulent stress for the *i*-phase. All values are averaged on the chosen characteristic interval by time. Index *m* belongs to the values at the axis of the flow (symmetry axis). The function-indicator of a phase in multiphase flow may be considered as the mathematical expectation, in contrast to the other multiphase approaches [8-10], which are based on introduction of the volumetric specific content of a phase in multiphase flow. But using the phase function-indicator allows computing the volumetric specific content of the phases introduced by another multiphase approaches.

The model for considered two-phase jet flow of immiscible liquids is presented in Fig.1 together with the schematic representation:

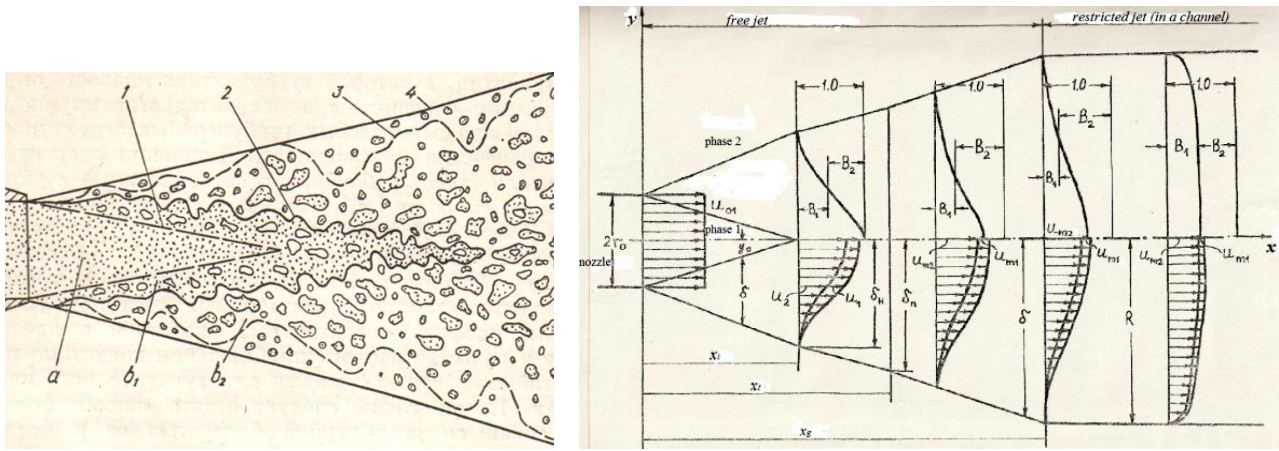

Fig. 1. General view of the heterogeneous turbulent jet and schematic representation:
$r_0$ – radius of a nozzle, $u_{01}$ - velocity at the nozzle; the conical surface 1 (in Fig. 1 to the left) is a boundary of a homogeneous potential core $a$, the internal sublayer $b_1$ contains as a disperse phase an ejected liquid, external sublayer $b_2$ contains as a disperse phase the liquid outgoing from the nozzle, the internal and external sublayers are divided by the phase inverse 2; and the surface 3 is a boundary dividing the turbulent and laminar zones 3, which the most indefinite one; 4 is an external conical surface of the axisymmetrical jet mixing zone (conditionally smooth)

Due to nonstationarity and instability, as well as the limited regularity of the processes occurring in turbulent jets, surfaces 2 and 3 in Fig. 1 to the left are blurred into the corresponding regions of inversion and intermittency. The external boundary of the jet is the outer envelope surface 4 of the set of surfaces 3. The first liquid from the nozzle is supposed with a uniform velocity profile. Surrounding second liquid (phase 2) is immovable before action of the first liquid going from the nozzle. The structural scheme for mixing of a jet with surrounding liquid according to Fig. 1 is simplified: the initial part of the length $x_i$ with the approximately linear boundaries for the conical surface (in cylindrical coordinate system) of the internal core of a first phase and mixing zone between internal and external boundaries of the jet. The mixing turbulent zone contains fragments of the phases as far as immiscible liquids have behaviors like the separate phases, with their interfacial multiple surfaces. After an initial part of the mixing zone, where a first phase in a potential core is totally spent, the short transit area follows. Then the ground part of the turbulent two-phase jet with the two phases well mixed across the entire layer of the two-phase jet.

Except the parameters of the phases, we have function-indicator of phase, which shows influence of each phase at each point of a space. Normally for description of multiphase flows the spatial averaging of the conservation equations of mass, impulse and energy is performed based on the concept of volumetric phase content [8-10], which does not fit so well to experimental study of a movement of the separate phases in a mixture. In a contrast to this, an approach [1] with its special experimental technology and micro sensor for the measurements in two-phase flows fits well for such flows. Actually all known methods of multiphase flows are well connected and the parameters averaged by time [1] can be easily transformed to the ones averaged by space [8-10]. The external interface of the mixing zone is determined by zero longitudinal velocity of the second phase and transversal velocity of the first phase (the second phase is sucked from immovable surrounding into the mixing zone). The function-indicator of the first phase $B_1(t)$ is zero at the external interface because the first phase is absent in surroundings. Similar, the function-indicator $B_2(t)$ is zero on the interface of the potential core, boundary of the first phase going from the nozzle. In a first approach, an influence of the mass, viscous and capillary forces is neglected. With account of the above-mentioned, the boundary conditions are [1]:

$$y=y_0, \quad u_i=u_{0i}, \quad v_i=0, \quad \tau_i=0, \quad B_1=1; \qquad y=y_0+\delta, \quad u_i=0, \quad v_i=0, \quad \tau_i=0, \quad B_1=0. \qquad (2)$$

## 2. Basic equations and polynomial profiles for the free turbulent heterogeneous jet

The turbulent stress in the phase is stated by the "new" Prandtl's formula $\tau_i = \rho_i \kappa_i \delta u_{mi} \partial u_i / \partial y$, where $\kappa_i$ is the coefficient of turbulent mixing for $i$-th phase, $\delta$ is the width of the mixing layer. The polynomial approximations for the velocity profiles and other functions in the turbulent mixing zone have been obtained based on the boundary conditions (2) [1, 2], e.g.

$$u_1 / u_{01} = 1 - 4\eta^3 + 3\eta^4, \quad u_2 / u_{02} = 1 - 6\eta^2 + 8\eta^3 - 3\eta^4, \tag{3}$$

$$B_1 = B_1^{(0)} = 1 - \eta^3 + 0.5\eta^2(1-\eta)h(x), \quad h \in [-6, 0], \tag{4}$$

where the function $h(x) = \left(\partial^2 B_1 / \partial \eta^2\right)_{\eta=0}$ is responsible for a transition of the piecewise-continuous function-indicator $B_1^{(n)}$ to its next approximation, determined from the condition that the derivative by $\eta$ with respect to a point $\eta = 1$ is equal zero up to $(n+1)$-th and including order. These functions have the breaks at the transition points of the permanent characteristic function $B_1^{(n)}(\eta, h)$ from the one regional approximation to the other one (a first derivative has break at those points). It is impossible to get a common approximation for the function $B_1(\eta, h)$ satisfying the boundary conditions in all range by parameter $i_0$ (due to requirement of variation $B_1$ from 0 to 1).

Based on the above considered, the integral correlations have been derived for the initial part of a jet [1, 2] integrating the momentum conservation equation for the total cross-section of a flow of the two-phase mixture for $y=y_0+\delta$ and $y=y^*$, respectively. The polynomial approximations for the functions $u_2$, $B_1$ on a ground part of a jet the same is kept, but for the function $u_1$ approximation is as follows

$$u_1 / u_{m1} = 1 - 3\eta^2 + 2\eta^3, \tag{5}$$

The integral correlations for the ground part of a jet obtained similarly to the initial part of a jet. Also, the momentum conservation equation for the total and for the part of the cross section, respectively, according to [11] was got. And the momentum equation on the jet's axis ($y=0$) is used too:

$$\sum_{j=1}^{2} \rho_j B_{mj} u_{mj} \frac{du_{mj}}{dx} = 2 \sum_{j=1}^{2} \left[ \frac{\partial}{\partial y} \left( B_j \tau_j \right) \right]_m. \tag{6}$$

The model equations are implemented for analysis and numerical simulation the basic features of the stationary turbulent two-phase jet of two immiscible liquids. The function-indicator $B_1$ shows how much is a presence of the first phase in a selected point of mixing zone, which can be directly compared to an experimental data by two-phase sensor [1]. Therefore, a solution of the task may give both parameters of the flow together with their belonging to a particular phase.

## 3. Dimensionless equation array for free two-phase jet

Dimensionless equation array with account of (1) - (6), for initial part of a free jet is as follows [1, 2, 7]:

$$y_0^2 + 2\delta \sum_{j=1}^{2} y_0^{2-j} \delta^{j-1} a_j = 1, \quad y_0^2 + 2\delta \sum_{j=1}^{2} y_0^{2-j} \delta^{j-1} \left( a_{j+2} + i_0 b_{j+2} \right) = 1, \tag{7}$$

$$\left(1 - u_1^*\right) y_0 \frac{dy_0}{d\varsigma} + \frac{d}{d\varsigma} \delta \sum_{j=1}^{2} y_0^{2-j} \delta^{j-1} \left( a_{j+2}^* + i_0 b_{j+2}^* \right) - \frac{d}{d\varsigma} \delta \sum_{j=1}^{2} y_0^{2-j} \delta^{j-1} \left( a_j^* u_1^* + i_0 b_j^* u_2^* \right) = \left( y_0 + \delta \eta^* \right) =$$

$$= \left(y_0 + \delta\eta^*\right)\sum_{j=1}^{2} B_j^* \left(\frac{\partial u_j}{\partial \eta}\right)^* \left(i_0 \kappa_{21}\right)^{j-1}, \qquad B_1 + B_2 = 1.$$

where:

$$\bar{y}_0 = y_0/r_0, \quad \bar{\delta} = \delta/r_0, \quad \eta = (y - y_0)/\delta, \quad \bar{x} = x/r_0, \quad \varsigma = \kappa_1 \bar{x}, \quad s_0 = u_{02}/u_{01}, \quad i_0 = ns_0^2,$$

$$n = \rho_2/\rho_1, \quad \kappa_{21} = \kappa_2/\kappa_1, \quad a_i = a_{i1} + a_{i2}h, \quad b_i = b_{i1} + b_{i2}h, \tag{8}$$

$$a_i = \int_0^1 B_1 \bar{u}_1 \eta^{j-1} d\eta, \quad b_i = \int_0^1 B_2 \bar{u}_2 \eta^{j-1} d\eta \ (i=1,2); \quad a_i = \int_0^1 B_1 \bar{u}_1^2 \eta^{j-1} d\eta, \quad b_i = \int_0^1 B_2 \bar{u}_2^2 \eta^{j-1} d\eta \ (i=3,4); \ j=1,2.$$

Except the above, for the dimensionless parameters the same notations are used further, as for the dimensional ones. Only in (8) it is stated for clarification. The sliding factor $s_0$ is assumed constant, the parameters at $\eta=\eta^*<1$ are signed with a star *. The system (8) must satisfy the following boundary conditions

$$\varsigma = 0, \ y_0 = 1, \ \delta = 0; \quad \varsigma = \varsigma_i, \ y_0 = 0, \ \delta = \delta_i; \tag{9}$$

where $\varsigma_i$, $\delta_i$ are the dimensionless length of a jet and its maximal radius (at the end of initial part). In dimensionless form the scales $r_0$, $\delta$, $u_{0i}$ are accepted for the longitudinal and transversal coordinates and velocity.

The functions $y_0(\varsigma)$, $\delta(\varsigma)$, $h(\varsigma)$ computed from (7), (9) allow obtaining the other characteristics for a jet by the stated parameters $i_0, \kappa_1, \kappa_2$. The main advantage of the model is a possibility to have all characteristics of a flow together with their belonging to a particular phase: functions $B_1$, $B_2$. The transversal velocities' distributions, interface interactions, the coefficients of the volumetric $q$ and mass ejection $g$ and kinetic energy $e_i$ for the phases in a flow are computed as follows:

$$\frac{B_1 v_1}{\kappa_1 u_{01}} = B_1 u_1 \frac{d}{d\varsigma}\left(y_0 + \delta\eta\right) - \frac{1}{y_0 + \delta\eta}\frac{d}{d\varsigma}\left(0.5 y_0^2 + a_1 y_0 \delta + a_2 \delta^2\right),$$

$$\frac{B_2 v_2}{\kappa_1 u_{02}} = B_2 u_2 \frac{d}{d\varsigma}\left(y_0 + \delta\eta\right) - \frac{1}{y_0 + \delta\eta}\frac{d}{d\varsigma}\left(b_1 y_0 \delta + b_2 \delta^2\right),$$

$$R_{21} = -\frac{1}{\left(y_0 + \delta\eta\right)\delta}\frac{\partial u_1}{\partial \eta}\left\{\frac{d}{d\varsigma}\left(0.5 y_0^2 + a_1 y_0 \delta + a_2 \delta^2\right) + \frac{\partial}{\partial \eta}\left[\left(y_0 + \delta\eta\right)B_1 \frac{\partial u_1}{\partial \eta}\right]\right\},$$

$$q = 2s_0 \delta\left(b_1 y_0 + b_2 \delta\right), \quad e_1 = y_0^2 + 2\int_0^1 B_1 \bar{u}_1^3 \left(y_0 + \delta\eta\right)\delta d\eta, \quad e_2 = 2i_0 s_0 \int_0^1 B_2 \bar{u}_2^3 \left(y_0 + \delta\eta\right)\delta d\eta.$$

The dimensionless equation array for the ground part of the turbulent two-phase jet is the next

$$2B_{m1} u_{m1} \delta^2 \sum_{j=1}^{2} \alpha_{1j} h^{j-1} = 1, \quad 2B_{m1} u_{m1} \delta^2 \sum_{j=1}^{2} \left(\alpha_{2j} + i_0 \beta_{2j}\right) h^{j-1} + i_0 \beta_{20} = 1, \tag{10}$$

$$\frac{d}{d\varsigma} u_{m1}^2 \delta^2 \left[B_{m1} \sum_{j=1}^{2}\left(\alpha_{2j}^* + i_0 \beta_{2j}^*\right)h^{j-1} + i_0 \beta_{20}^*\right] - u_{m1} \frac{d}{d\varsigma}\left(u_{m1} \delta^2\right)\left[B_{m1} \sum_{j=1}^{2}\left(\alpha_{1j}^* u_1^* + i_0 \beta_{1j}^* u_2^*\right)h^{j-1} + i_0 u_2^* \beta_{10}^*\right] =$$

$$= \eta^* \delta u_{m1}^2 \left[\left(1 - i_0 \kappa_{21}\right)B_{m1} \sum_{j=1}^{2}\left(\frac{\partial u_j}{\partial \eta}\right)^* \gamma_j^* h^{j-1} + i_0 \kappa_{21}\left(\frac{\partial u_2}{\partial \eta}\right)^*\right].$$

Here are:

$$\bar{x} = \frac{x - x_t}{r_0}, \quad \varsigma = \kappa_1 \bar{x}, \quad \bar{u}_{mi} = \frac{u_{mi}}{u_{0i}}, \quad \bar{u}_i = \frac{u_i}{u_{mi}}, \quad i_0 = ns_0^2, \quad \bar{B}_2 = \frac{B_2}{B_{m1}}, \quad \bar{B}_1 = \frac{B_1}{B_{m1}} = \gamma_1 + \gamma_2 h,$$

$$\int_0^1 \bar{B}_1 \bar{u}_1^i \eta\, d\eta = \sum_{j=1}^{2} \alpha_{ij} h^{j-1}, \quad \int_0^1 \bar{B}_2 \bar{u}_2^i \eta\, d\eta = \frac{\beta_{i0}}{B_{m1}} + \beta_{i1} + \beta_{i2} h \quad (i=1, 2). \qquad (11)$$

In all further equations we keep previous assignments for dimensionless parameters as for the dimension ones. Star * means a value by $\eta=\eta^*<1$, $x_t$ is the end of a jet's transient part. It is assumed $u_{m2}=s_0 u_{m1}$ ($s_0$=const), which means that sliding of the phases is preserved the same as for the initial part of a jet. The boundary condition for the equation array (16) are

$$\zeta=0, \quad u_{m1}=1, \quad B_{m1}=1, \quad \delta=\delta_t; \quad \zeta=\infty, \quad u_{m1}=0, \quad B_{m1}=0, \quad \delta=\infty; \qquad (12)$$

$\delta_t$ is a radius of the jet at the transient cross section. Solution of the boundary problem for differential equation array (10) with boundary conditions (12) yields the functions $u_{m1}(\zeta)$, $B_{m1}(\zeta)$, $\delta(\zeta)$ and $h(\zeta)$ for the stated parameters $i_0$, $\kappa_{21}$. Similarly, for the initial part of a jet. Afterward, also the distribution of all parameters of a two-phase flow is calculated: turbulent stresses, mass flow rate, trajectories of the phases in a mixing layer.

## 4. Analytical-numerical solution of the equations for a free two-phase jet

The combined analytical-numerical solution of the boundary problems for the mixed equation arrays (7), (9) and (10), (12) was obtained and analyzed in a range of varying parameters [1, 2, 7, 12]. The correspondence with experimental data was revealed quite well [1], as seen from the Fig. 2:

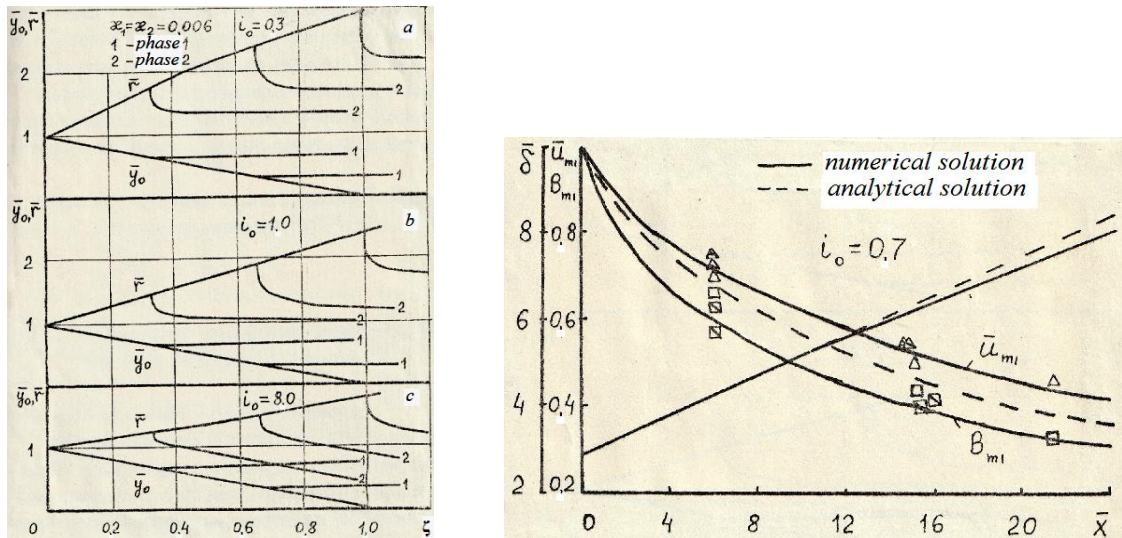

Fig. 2 The stream lines on initial part and characteristic of ground part in a turbulent two-phase jet

On the initial part there are the following peculiarities. Increase of density leads to shortening the length of an initial part. The profiles are close to the auto model ones. The maximal value of a turbulent stress of a flow with increase of the density of ejected liquid is shifted to an external boundary of the mixing layer. The profiles $B_1$, $B_1 \bar{u}_1$, $\overline{\langle \rho u^2 \rangle}$ become more full, while $B_2 \bar{u}_2$ is exhausting. The transversal velocity of the first phase is mostly directed to the external boundary of the mixing layer, while the second phase is moving in an opposite direction, to the axis of a jet. This tendency is the clearest by big values of $i_0$. The trajectories of phases in a mixing layer in Fig. 2 to the left show that the first liquid keeps a tendency to flow parallel to a jet's axis for all density ratios, while the second phase is ejected into a mixing layer more and more intensively with increase of its density. The real intensive mixing is seen by high density ratio ($i_0 = n s_0^2 =8$) but as far as this parameter $i_0$ contains substantial influence of a slip of phases $s_0 = u_{02}/u_{01}$, which is falling down with increase of the density ratio, then $i_0=8$ may correspond to a density ratio even higher than 8. The initial and the ground parts of a jet must be connected. For this, the length of a transient part of a jet was determined by interpolation of the external boundary of a jet from the initial to the ground part. This approximation was: $\delta = 1 + \alpha_1 \varsigma + \alpha_2 \varsigma^2$, where from the value $x_t$ was computed by the known $\kappa_1, \delta_i$. Value $\kappa_1$ was taken equal to the corresponding value at the initial part. The coefficients are: $i_0=0.3$, $\alpha_1=56$, $\alpha_2=-290$; $i_0=1.0$, $\alpha_1=57.5$, $\alpha_2=-250$; $i_0=8.0$, $\alpha_1=29$, $\alpha_2=600$.

An interesting feature was revealed by influence of the parameters $i_0$, $\kappa_{21}$ on a solution. The radius of a jet and velocities of phases on the axis practically don't depend on $\kappa_{21}$ being totally determined by the value $i_0$, while the functions $B_{m1}$ and $h$ strictly depend on $\kappa_{21}$. Thus, the turbulent mixing influences mostly the internal structure of a flow, the distribution of phases and velocities depend on density ratio of the phases (internal structure has little influence on it). The velocity distribution, in a turn, determines the radius of a mixing zone because it changes with falling of the velocity according to the mass and momentum conservation equations. With a growth of the ratio $\kappa_{21}$ the expansion of the jet occurs more intensively. The turbulent stress is maximal approximately at the distance of about 1/3 of the mixing layer.

The transversal velocity of the first phase (ejecting liquid) $B_1 v_1$ is substantially lower than the transversal velocity of the second phase (ejected liquid) $B_2 v_2$, which is determining a transversal flow in a mixing layer. Independent of the value $i_0$ the velocity of the second phase is directed to an axis and achieves the maximum on a distance about $0.5\delta$, decreasing with the growth of the density of ejected liquid. The results obtained have been compared to the experimental data of a number of researchers [1, 11, 14, 15]. The correspondence was good for the constants as follows: $i_0=0.3$, $\kappa_{21}=0.6$; $i_0=1$, $\kappa_{21}=0.4$; $i_0=8.0$, $\kappa_{21}=0.2$. The comparison with experimental data for $i_0=0.7$ (oil-water) using the special two-phase sensor [1] was the best at $\kappa_{21}=0.46$. The stream lines in Fig. 2 are presented in relation to the length of the initial part ($\zeta/\zeta_i$), therefore shortening of the initial part of a jet with increase of a density of the ejected liquid is already accounted.

Because $\delta(\zeta, i_0)$ are close to the linear functions, an approximation $\delta(\zeta, i_0) = c_1 + c_2\zeta$ was used. For $\kappa_{21}=0.5$, the first equation of the system (10) is simplified and solved analytically:

$$\frac{du_{m1}}{d\varsigma} = -12\frac{u_{m1}}{c_1 + c_2\varsigma},$$

which is the more precise, the closer $\kappa_{21}$ is to 0.5. From the above-considered, it corresponds to $i_0<1$. The solution, with account of the (12) and $\zeta = \infty$, $u_{m1}\delta = const$, $u_{m1}\delta^2 = \infty$, is as follows

$$u_{m1} = \delta_t / (\delta_t + 12\zeta), \quad \delta = \delta_t + 12\zeta. \tag{13}$$

The obtained results can be used in engineering computations and estimations, e.g. the correlation (13) for the radius of the mixing layer and velocity at the axis, as well as function-indicator of phases.

## 5. The phenomenon of amplification of a jet power relation to the input power

As shown in Fig. 3 for the two-phase jet power $\langle \overline{\rho u^2} \rangle (\eta)$ for 3 values of $i_0=0.3$; 1.0; 8.0, there is exciting phenomenon of amplification of jet power relation to the input power of the jet of first phase at its exit from the nozzle. This effect reveals only for $i_0 \gg 1$, e.g. $i_0=8.0$. First, we discovered this new phenomenon at the end of 1970-th [1, 2]. We could not prove it experimentally due to serious impediments in measuring under such conditions. What is more, it is inside the usual range of inaccuracy in turbulent flow experiments. Therefore, we have just carefully proved it a few times theoretically to be sure.

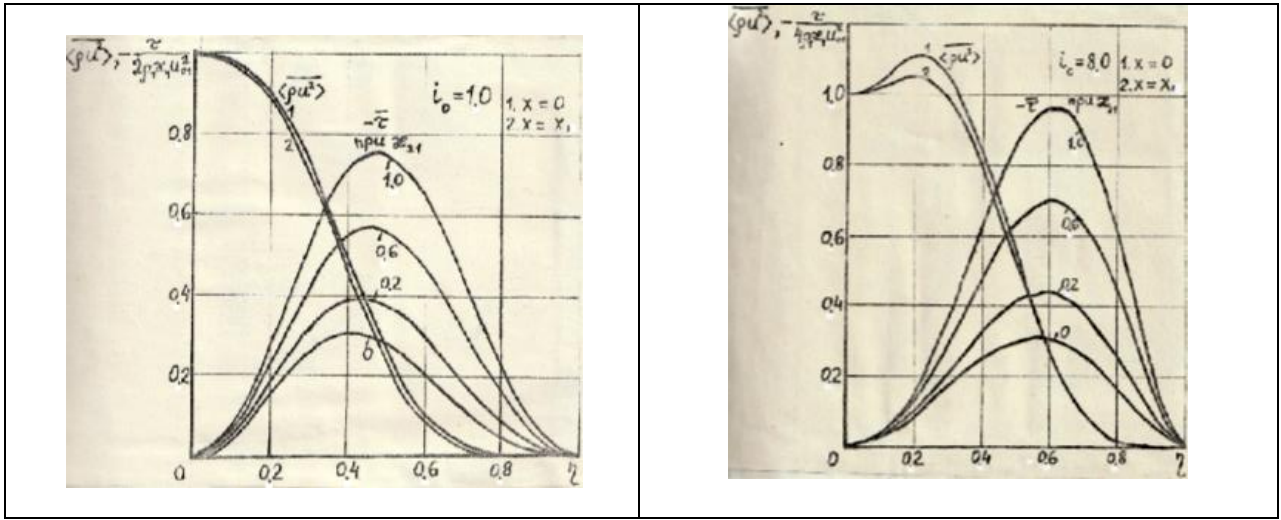

Fig. 3. Functions $\left\langle \overline{\rho u^2} \right\rangle(\eta)$ for $i_0$=1.0; 8.0

Later on, we have met a few reports in a literature, mainly from the gas flows in the rocket engines [16, 17]. The causes of this effect and the source of additional energy are considered by Kotousov S.L. [16] who considered a number of available mechanicsms, e.g. decided in his investigation that the acceleration of the jet is caused not only by the inlet pressure, but also by the decrease in the potential energy of the current medium by reducing its absolute pressure down to the level of the technical vacuum. It follows that the effect studied by Kotousov can not only explain the mechanism of the operation of various hydrodynamic constructions (Schauberger, Clem, Cheryu, etc.), but also has tremendous practical value in itself. This is an evidence that simple model of turbulent jets like we use here is far from reality in many ascpects, though in general is in good correspondence with experimental data.

Our explanation is as follows. Due to ejection of the second phase at the nozzle, close to it some depressurization occurs, which cause some additional amount of the second phase involved in mixing flow. As seen from Fig.3, for $i_0$=8.0, about 5% of the total jet power is added to the input power of the jet outgoing from the nozzle. At the end of the initial part of the jet the corresponding value is about 2.5%, so that nearly twice less than at the nozzle cross section. As concern to high density ratio ($i_0$=8.0), it is probably explained by big energy of the small additional amount of a second phase due to its high density (by the same velocity the denser liquid has higher power).

## 6. Basic equations for heterogeneous turbulent two-phase jet in a confined space

The free jet can be confined at its initial or ground part, so that depending on it different physical situation take place. Let us start with the first model shown in Fig. 4. This flow is described by the equation array (3). In a potential core the Bernoulli equation must be satisfied

$$p + 0.5\rho_1 u_{c1}^2 = p_0 + 0.5\rho_1 u_{01}^2,  \qquad (14)$$

where $u_{ci}$ is velocity of $i$-phase in a core of a jet, $x_c$ is the initial cross-section of the confining a jet chamber, $p_0$ is a pressure in a free jet. From (14) follows

$$u_{c1} = u_{01}\sqrt{1 - \bar{p}}, \quad \bar{p} = 2\frac{p - p_0}{\rho_1 u_{01}^2}. \qquad (15)$$

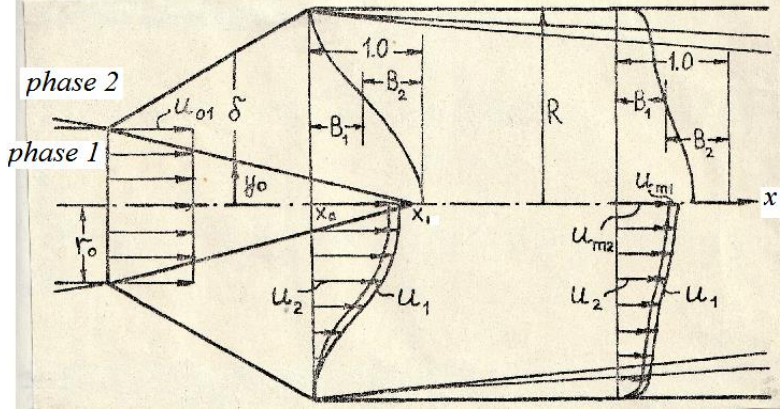

Fig. 4. Turbulent two-phase jet confined on its initial part

Using the transversal velocities of the phases from the first two equations of (1), with account of zero transversal velocities at the axis due to symmetry of a flow and at the wall of a chamber due to impermeability, we can get

$$\rho_1 u_{01} r_0^2 = \rho_1 u_{c1} y_0^2 + 2 \int_{y_0}^{R} \rho_1 B_1 u_1 (y_0 + y) dy, \quad \int_{y_{0c}}^{R_c} \rho_2 B_{2c} u_{2c} (y_{0c} + y) dy = \int_{y_0}^{R} \rho_2 B_2 u_2 (y_0 + y) dy,$$

$$0.5 \frac{d}{dx}\left(\rho_1 u_{c1}^2 y_0^2\right) + \frac{d}{dx} \sum_{i=1}^{2} \int_{y_0}^{R} \rho_i B_i u_i^2 (y_0 + y) dy = R \sum_{i=1}^{2} B_{wi} \tau_{wi} - 0.5 R^2 \frac{dp}{dx}, \qquad (16)$$

$$0.5 \frac{d}{dx}\left[\rho_1 u_{c1}(u_{c1} - u_1^*) y_0^2\right] + \frac{d}{dx} \sum_{i=1}^{2} \int_{y_0}^{y^*} \rho_i B_i u_i^2 (y_0 + y) dy - \sum_{i=1}^{2} u_i^* \frac{d}{dx} \int_{y_0}^{y^*} \rho_i B_i u_i (y_0 + y) dy =$$

$$= (y_0 + y^*) \sum_{i=1}^{2} B_i^* \tau_i^* - 0.5(y_0 + y^*)^2 \frac{dp}{dx},$$

where $R(x)$ is the radius of the confining chamber, which is equal to the radius of the free jet at the cross section $x=x_c$. Indexes $c$ and $w$ belong to $x=x_c$ and to the wall of a chamber, respectively. Star assigns the values at $y=y^*<\delta$, e.g. $y^*=0.5\delta$. The third integral correlation in (16) was derived with account of the following two equations got from the first two equations of (16):

$$\frac{d}{dx} \int_{y_0}^{R} \rho_1 B_1 u_1 (y_0 + y) dy = -\frac{1}{2} \frac{d}{dx} \rho_1 u_{c1} y_0^2, \quad \frac{d}{dx} \int_{y_0}^{R} \rho_2 B_2 u_2 (y_0 + y) dy = 0. \qquad (17)$$

The equation array (16) contains two equation of the mass conservation for the first and second phase and the integral correlations of momentum – by total cross-section and by part of it. Correspondingly, 2 algebraic and 2 ordinary differential equations. It is model of the confined two-phase jet for $x \geq x_c$, where by $x \geq x_i$, $y_0=0$, and the function-indicator of the first phase is not constant at the axis of a jet (chamber): $B_{m1}=B_{m1}(x)$. The equations (16) satisfy the following boundary conditions

$$x = x_c, \quad y_0 = y_{0c}, \quad p=p_0, \quad u_i = u_{ic}, \quad B_1 = B_{1c}. \qquad (18)$$

If similarly to the free jet, an assumption about the relation

$$u_{2c}/u_{1c} = u_{m2}/u_{m1} = s_0 = \text{const}, \qquad (19)$$

is accepted, then the system (16) is reduced to the following dimensionless form:

$$\sqrt{1-\bar{p}}\left[\bar{y}_0^2 + 2\bar{\delta}\int_0^1 B_1\bar{u}_1(\bar{y}_0+\bar{\delta}\eta)d\eta\right] = 1,$$

$$\sqrt{1-\bar{p}}\int_0^1 (1-B_1)\bar{u}_2(\bar{y}_0+\bar{\delta}\eta)\bar{\delta}d\eta = \int_0^1 (1-B_{1c})\bar{u}_{2c}(\bar{y}_{0c}+\bar{\delta}_c\eta)\bar{\delta}_c d\eta,$$

$$0.5\frac{d}{d\bar{x}}\bar{y}_0^2(1-\bar{p}) + \frac{d}{d\bar{x}}\sum_{i=1}^2 i_0^{i-1}(1-\bar{p})\int_0^1 B_i\bar{u}_i^2(\bar{y}_0+\bar{\delta}\eta)\bar{\delta}d\eta = -\frac{\bar{R}^2}{4}\frac{d\bar{p}}{d\bar{x}} + \bar{R}\sum_{i=1}^2 B_i\bar{\tau}_i,$$

$$0.5(1-\bar{u}_1^*)\frac{d}{d\bar{x}}\bar{y}_0^2(1-\bar{p}) + \frac{d}{d\bar{x}}\sum_{i=1}^2 i_0^{i-1}(1-\bar{p})\int_0^{\eta^*} B_i\bar{u}_i^2(\bar{y}_0+\bar{\delta}\eta)\bar{\delta}d\eta + \quad (20)$$

$$-\sqrt{1-\bar{p}}\sum_{i=1}^2 i_0^{i-1}\bar{u}_i^*\frac{d}{d\bar{x}}\sqrt{1-\bar{p}}\int_0^{\eta^*} B_i\bar{u}_i(\bar{y}_0+\bar{\delta}\eta)\bar{\delta}d\eta = -\frac{(\bar{y}_0+\bar{\delta}\eta^*)^2}{4}\frac{d\bar{p}}{d\bar{x}} + (\bar{y}_0+\bar{\delta}\eta^*)\sum_{i=1}^2 B_i^*\bar{\tau}_i^*.$$

where: $\bar{\delta}=\dfrac{\delta}{r_0}$, $\eta=\dfrac{y-y_0}{\delta}$, $\bar{x}=\dfrac{x-x_c}{r_0}$, $\bar{y}_0=\dfrac{y_0}{r_0}$, $\bar{R}=\dfrac{R}{r_0}$, $n=\dfrac{\rho_2}{\rho_1}$, $\bar{u}_i=\dfrac{u_i}{u_{ci}s^{i-1}}$, $\bar{\tau}_i=\dfrac{\tau_i}{\rho_1 u_{01}^2}$, $i_0=ns_0^2$. At the $x>x_i$, $u_i=\bar{u}_i u_{mi}$. The boundary conditions are:

$$\bar{x}=0, \quad \bar{u}_i=\bar{u}_{ic}, \quad B_i=B_{ic}, \quad \bar{y}_0=\bar{y}_{0c}, \quad \bar{\delta}=\bar{\delta}_c, \quad \bar{p}=0. \quad (21)$$

At the initial confined part of a jet, $x\in(x_c, x_i)$, there is added also

$$\eta=0, \quad \bar{u}_i=1, \quad B_{m1}=1. \quad (22)$$

After disappearance of the potential core (in case a jet is confined at its initial part), $x>x_i$, the regularities of a jet's spreading are described by the following mixed system of the algebraic and differential equations:

$$\bar{\delta}=\bar{R}, \quad \bar{y}_0=0, \quad 2\bar{R}^2\sqrt{1-\bar{p}}\int_0^1 B_1\bar{u}_1\eta d\eta=1, \quad \frac{d}{d\bar{x}}(1-\bar{p})\bar{R}^2\sum_{i=1}^2 i_0^{i-1}\int_0^1 B_i\bar{u}_i^2\eta d\eta = \bar{R}\sum_{i=1}^2 B_{wi}\bar{\tau}_{wi} - \frac{\bar{R}^2}{4}\frac{d\bar{p}}{d\bar{x}},$$

$$\bar{R}^2\sqrt{1-\bar{p}}\int_0^1 (1-B_1)\bar{u}_2\eta d\eta = \int_0^1 (1-B_{1c})\bar{u}_{2c}(\bar{y}_{0c}+\bar{\delta}_c\eta)\bar{\delta}_c d\eta, \quad (23)$$

$$\frac{d}{d\bar{x}}(1-\bar{p})\bar{R}^2\sum_{i=1}^2 i_0^{i-1}\int_0^{\eta^*} B_i\bar{u}_i^2\eta d\eta - \sqrt{1-\bar{p}}\sum_{i=1}^2 i_0^{i-1}\bar{u}_i^*\frac{d}{d\bar{x}}\bar{R}^2\sqrt{1-\bar{p}}\int_0^{\eta^*} B_i\bar{u}_i\eta d\eta =$$

$$= \bar{R}\eta^*\sum_{i=1}^2 B_i^*\bar{\tau}_i^* - \frac{1}{4}\bar{R}^2(\eta^*)^2\frac{d\bar{p}}{d\bar{x}}.$$

In case of confining a jet at its ground part, the basic equations are derived similarly in dimension form:

$$\int_0^{R(x)} B_i u_i y\,dy = \int_0^{R_c} B_{ic}u_{ic}y\,dy, \quad (i=1,2), \qquad \sum_{i=1}^2 \rho_i B_{mi}\frac{du_{mi}}{dx} = 2\left(\frac{\partial}{\partial y}\sum_{i=1}^2 B_i\tau_i\right)_m - \frac{dp}{dx},$$

$$\frac{d}{dx}\sum_{i=1}^2\int_0^{y^*(x)}\rho_i B_i u_i^2 y\,dy - \sum_{i=1}^2 u_i^*\frac{d}{dx}\int_0^{y^*(x)}\rho_i B_i u_i y\,dy = y^*\sum_{i=1}^2 B_i^*\tau_i^* - 0.5(y^*)^2\frac{dp}{dx}, \quad (24)$$

$$\frac{d}{dx}\sum_{i=1}^2\int_0^{R(x)}\rho_i B_i u_i^2 y\,dy = -0.5R^2(x)\frac{dp}{dx} + R(x)\sum_{i=1}^2 B_{wi}\tau_{wi}.$$

## 7. Methodology of modeling of the confined two-phase jets

Further investigation of the stationary 2-D axisymmetrical heterogeneous two-phase turbulent jets in a confined domain is reasonable to conduct on some simplified models to reveal the basic features of such flows. First of all, the main problem with multiphase flow consists in closeness of the equation array. For the turbulent mixing coefficient, as before in the free turbulent jets, we use the "new" Prandtl formula. The peculiarity of the confined jet is an appearance of the boundary layer near the wall of a chamber [18], which complicates the structure of a flow substantially. If we assume a jet character of a flow inside a chamber up to the wall of the chamber, the both phases have slip on the wall, and consider the structural scheme similar to the proposed in for homogeneous jet [18], then $\eta^*$ is chosen the one corresponding to a developed jet flow. In such case, the profiles of the parameters are

$$u_i = (u_{mi} - u_{wi})\bar{u}_{ic} + u_{wi}, \quad (i=1,2), \quad B_1 = (B_{m1} - B_{w1})\bar{B}_1 + B_{w1}, \quad B_2 = 1 - B_1, \qquad (25)$$

where $\bar{u}_{1c}$, $\bar{u}_{2c}$, $\bar{B}_1$ are determined according to the above-considered method of two-phase turbulent flows.

The model allows considering mutual interaction of the two-phase turbulent jet with confining chamber. The main parameters of the task are: $u_{m1}(x)$, $u_{w1}(x)$, $B_{m1}(x)$, $B_{w1}(x)$, $p(x)$ and $R(x)$, one of which can be stated as the optimal in a desired way. For example, for the grading of an gradientless stabilization chamber, in case of jet's confining at its ground part, $dp/dx=0$ is put in (24). Then the functions $u_{m1}(x)$, $u_{w1}(x)$, $B_{m1}(x)$, $B_{w1}(x)$ and $R(x)$ are determined according to the task stated. For the total stabilization of a flow, when $u_{m1} = u_{w1} = u_{1\infty}$, $B_{m1} = B_{w1} = B_{1\infty}$, the derived mathematical model yields:

$$\bar{R}^2 \bar{B}_{1\infty} \bar{u}_{1\infty} = 2\int_0^1 \bar{B}_{1c} \bar{u}_{1c} \eta d\eta, \quad \bar{R}^2 \bar{B}_{1\infty} \bar{u}_{1\infty} = 2\int_0^1 \bar{B}_{1c} \bar{u}_{2c} \eta d\eta + 1/B_{m1c}\left(\bar{R}^2 \bar{u}_{1\infty} - 2\int_0^1 \bar{u}_{2c} \eta d\eta\right), \quad (i=1,2),$$

$$\frac{d\bar{p}}{dx} = -B_{m1c}\left[\bar{B}_{1\infty}(1-i_0) + i_0 / B_{m1c}\right]\frac{d\bar{u}_{1\infty}^2}{dx}. \qquad (26)$$

From (26) follows that $\bar{R}^2 \bar{u}_{1\infty}$ independently on $\bar{R}, \bar{B}_{1\infty}$ is determined by

$$C_0 = \bar{R}^2 \bar{u}_{1\infty} = 2\left[B_{m1c}\left(\int_0^1 \bar{B}_{1c}\bar{u}_{1c}\eta d\eta - \int_0^1 \bar{B}_{1c}\bar{u}_{2c}\eta d\eta\right) + \int_0^1 \bar{u}_{2c}\eta d\eta\right]. \qquad (27)$$

And pressure correlation with a form of the confining chamber is as follows

$$\bar{p} + B_{m1c}\left[\bar{B}_{1\infty}(1-i_0) + i_0 / B_{m1c}\right]C_0^2 / \bar{R}^4 = const, \qquad (28)$$

where $R = \bar{R}R_c$, $u_{1\infty} = \bar{u}_{1\infty}u_{m1c}$, index $\infty$ means a value of a parameter for totally stabilized flow, $\tau_{wi} = 0$ according to the accepted scheme.

If under the assumptions made, a flow of turbulent two-phase jet confined at its initial part, with a simplification $h = h_c = const$, is described by the following equations. For a jet's initial part ($B_{m1}=1$, $u_{mi} = u_{c1}s_0^{i-1}$):

$$2\sqrt{1-\bar{p}}(\bar{R}-\bar{y}_0)\int_0^1\left[(1-B_{w1})\bar{B}_{1c} + B_{w1}\right]\left[(1-\bar{u}_{w1})\bar{u}_{1c} + \bar{u}_{w1}\right]\left[(\bar{R}-\bar{y}_0)\eta + \bar{y}_0\right]d\eta = 1 - \bar{y}_0^2\sqrt{1-\bar{p}},$$

$$\sqrt{1-\bar{p}}(\bar{R}-\bar{y}_0)\int_0^1\left[(1-B_{w1}) + (B_{w1}-1)\bar{B}_{1c}\right]\left[(1-\bar{u}_{w1})\bar{u}_{2c} + \bar{u}_{w1}\right]\left[(\bar{R}-\bar{y}_0)\eta + \bar{y}_0\right]d\eta =$$

$$(\bar{R}_c - \bar{y}_{0c})\int_0^1(1-\bar{B}_{1c})\bar{u}_{2c}\left[(\bar{R}_c - \bar{y}_{0c})\eta + \bar{y}_{0c}\right]d\eta,$$

$$0.5\bar{y}_0^2\left(\bar{R}-\bar{y}_0\right)\int_0^1\left[\left(1-B_{w1}\right)\bar{B}_{1c}+B_{w1}\right]\left[\left(1-\bar{u}_{w1}\right)\bar{u}_{1c}+\bar{u}_{w1}\right]^2\left[\left(\bar{R}-\bar{y}_0\right)\eta+\bar{y}_0\right]d\eta+$$

$$+i_0\left(\bar{R}-\bar{y}_0\right)\int_0^1\left[\left(1-B_{w1}\right)+\left(B_{w1}-1\right)\bar{B}_{1c}\right]\left[\left(1-\bar{u}_{w1}\right)\bar{u}_{2c}+\bar{u}_{w1}\right]^2\left[\left(\bar{R}-\bar{y}_0\right)\eta+\bar{y}_0\right]d\eta=$$

$$=\frac{1}{\sqrt{1-\bar{p}}}\left\{\frac{\bar{y}_{0c}^2}{2}+\left(\bar{R}_c-\bar{y}_{0c}\right)\int_0^1\left[\bar{B}_{1c}\bar{u}_{1c}^2+i_0\left(1-\bar{B}_{1c}\right)\bar{u}_{2c}^2\right]\left[\left(\bar{R}_c-\bar{y}_{0c}\right)\eta+\bar{y}_{0c}\right]d\eta-\bar{R}^2\frac{\bar{p}}{4}\right\},$$

$$0.5\left[\left(1-\bar{u}_{w1}\right)+\left(\bar{u}_{w1}-1\right)\bar{u}_{1c}^*\right]d/d\zeta\left[\left(1-\bar{p}\right)\bar{y}_0^2\right]+ \qquad (29)$$

$$\frac{d}{d\zeta}\left(1-\bar{p}\right)\left\{\left(\bar{R}-\bar{y}_0\right)\int_0^{\eta^*}\left[\left(1-B_{w1}\right)\bar{B}_{1c}+B_{w1}\right]\left[\left(1-\bar{u}_{w1}\right)\bar{u}_{1c}+\bar{u}_{w1}\right]^2\left[\left(\bar{R}-\bar{y}_0\right)\eta+\bar{y}_0\right]d\eta+$$

$$+i_0\left(\bar{R}-\bar{y}_0\right)\int_0^{\eta^*}\left[\left(1-B_{w1}\right)+\left(B_{w1}-1\right)\bar{B}_{1c}\right]\left[\left(1-\bar{u}_{w1}\right)\bar{u}_{2c}+\bar{u}_{w1}\right]^2\left[\left(\bar{R}-\bar{y}_0\right)\eta+\bar{y}_0\right]d\eta\right\}-\sqrt{1-\bar{p}}\cdot$$

$$\cdot\left[\left(1-\bar{u}_{w1}\right)\bar{u}_{1c}^*+\bar{u}_{w1}\right]\frac{d}{d\zeta}\left(\bar{R}-\bar{y}_0\right)\int_0^{\eta^*}\left[\left(1-B_{w1}\right)\bar{B}_{1c}+B_{w1}\right]\left[\left(1-\bar{u}_{w1}\right)\bar{u}_{1c}+\bar{u}_{w1}\right]\left[\left(\bar{R}-\bar{y}_0\right)\eta+\bar{y}_0\right]\sqrt{1-\bar{p}}d\eta+$$

$$-\sqrt{1-\bar{p}}i_0\left[\left(1-\bar{u}_{w1}\right)\bar{u}_{2c}^*+\bar{u}_{w1}\right]\frac{d}{d\zeta}\sqrt{1-\bar{p}}\left(\bar{R}-\bar{y}_0\right)\int_0^{\eta^*}\left[\left(1-B_{w1}\right)+\left(B_{w1}-1\right)\bar{B}_{1c}\right]\left[\left(1-\bar{u}_{w1}\right)\bar{u}_{2c}+\bar{u}_{w1}\right]\cdot$$

$$\cdot\left[\left(\bar{R}-\bar{y}_0\right)\eta+\bar{y}_0\right]d\eta=\left(1-\bar{p}\right)\left[\left(\bar{R}-\bar{y}_0\right)\eta^*+\bar{y}_0\right]\left(1-\bar{u}_{w1}\right)^2\left\{\left[\left(1-B_{w1}\right)\bar{B}_{1c}^*+B_{w1}\right]\left(d\bar{u}_{1c}/d\zeta\right)^*+$$

$$i_0\kappa_{21}\left[\left(1-B_{w1}\right)+\left(B_{w1}-1\right)\bar{B}_{1c}^*\right]\left(d\bar{u}_{2c}/d\zeta\right)^*-0.25\left[\left(\bar{R}-\bar{y}_0\right)\eta+\bar{y}_0\right]^2 d\bar{p}/d\zeta.$$

The boundary conditions for the equation array (29) are stated as follows:

$$\zeta=0,\ y=y_{0c},\ h=h_c,\ B_{w1}=0,\ \bar{B}_{1c}=B_{1c},\ \bar{R}=\bar{R}_c,\ \bar{u}_{w1}=0,\ \bar{p}=0. \qquad (30)$$

Here are: $\zeta=\kappa_1\bar{x}$, $\kappa_{21}=\kappa_2/\kappa_1$.

By $x\geq x_i$, also the momentum equation at the axis is added, and the equation array is the next:

$$2\sqrt{1-\bar{p}}\bar{R}^2\int_0^1\left[\left(B_{m1}-B_{w1}\right)\bar{B}_{1i}+B_{w1}\right]\left[\left(1-\bar{u}_{w1}\right)\bar{u}_{1c}+\bar{u}_{w1}\right]\eta d\eta=1,$$

$$\sqrt{1-\bar{p}}\bar{R}^2\int_0^1\left[\left(B_{w1}-B_{m1}\right)\bar{B}_{1i}+1-B_{w1}\right]\left[\left(1-\bar{u}_{w1}\right)\bar{u}_{2c}+\bar{u}_{w1}\right]\eta d\eta=\bar{R}_i^2\int_0^1\left(1-B_{1i}\right)\bar{u}_{2c}\eta d\eta,$$

$$\int_0^1\left[\left(B_{m1}-B_{w1}\right)\bar{B}_{1i}+B_{w1}\right]\left[\left(1-\bar{u}_{w1}\right)\bar{u}_{1c}+\bar{u}_{w1}\right]^2\eta d\eta+$$

$$+i_0\int_0^1\left[\left(1-B_{w1}\right)+\left(B_{w1}-B_{m1}\right)\bar{B}_{1i}\right]\left[\left(1-\bar{u}_{w1}\right)\bar{u}_{2c}+\bar{u}_{w1}\right]^2\eta d\eta=$$

$$=\left\{\int_0^1\left[B_{wli}+\left(1-B_{wli}\right)\bar{B}_{1i}\right]\left[\left(1-\bar{u}_{wli}\right)\bar{u}_{1c}+\bar{u}_{wli}\right]^2\eta d\eta+ \right. \qquad (31)$$

$$\left.+i_0\int_0^1\left[\left(1-B_{wli}\right)+\left(B_{wli}-1\right)\bar{B}_{1i}\right]\left[\left(1-\bar{u}_{wli}\right)\bar{u}_{2c}+\bar{u}_{wli}\right]^2\eta d\eta=\frac{\bar{p}_i-\bar{p}}{4}\right\}\frac{\bar{R}_i^2}{\left(1-\bar{p}\right)\bar{R}^2},$$

$$\frac{d}{d\zeta}\bar{R}^2(1-\bar{p})\left\{(-\bar{y}_0)\int_0^{\eta^*}\left[(B_{m1}-B_{w1})\bar{B}_{1i}+B_{w1}\right]\left[(1-\bar{u}_{w1})\bar{u}_{1c}+\bar{u}_{w1}\right]^2\eta d\eta+\right.$$

$$\left.+i_0\int_0^{\eta^*}\left[(1-B_{w1})+(B_{w1}-B_{m1})\bar{B}_{1i}\right]\left[(1-\bar{u}_{w1})\bar{u}_{2c}+\bar{u}_{w1}\right]^2\eta d\eta\right\}+$$

$$-\sqrt{1-\bar{p}}\left[(1-\bar{u}_{w1})\bar{u}_{1c}^*+\bar{u}_{w1}\right]\frac{d}{d\zeta}\bar{R}^2\sqrt{1-\bar{p}}\int_0^{\eta^*}\left[(B_{m1}-B_{w1})\bar{B}_{1i}+B_{w1}\right]\left[(1-\bar{u}_{w1})\bar{u}_{1c}+\bar{u}_{w1}\right]\eta d\eta+$$

$$-i_0\sqrt{1-\bar{p}}\left[(1-\bar{u}_{w1})\bar{u}_{2c}^*+\bar{u}_{w1}\right]\frac{d}{d\zeta}\sqrt{1-\bar{p}}\bar{R}^2\int_0^{\eta^*}\left[(B_{m1}-B_{w1})\bar{B}_{1i}+(1-B_{w1})\right]\left[(1-\bar{u}_{w1})\bar{u}_{2c}+\bar{u}_{w1}\right]\eta d\eta=$$

$$=(1-\bar{p})\bar{R}^2\eta^*(1-\bar{u}_{w1})^2\left\{\left[(B_{m1}-B_{w1})\bar{B}_{1i}^*+B_{w1}\right](d\bar{u}_{1c}/d\eta)^*+\right.$$

$$\left.+i_0\kappa_{21}\left[(1-B_{w1})+(B_{w1}-B_{m1})\bar{B}_{1i}^*\right](d\bar{u}_{2c}/d\eta)^*\right\}-0.25(\eta^*)^2\bar{R}^2 d\bar{p}/d\zeta,$$

$$\bar{u}_{m1}\left[B_{m1}+i_0(1-B_{m1})\right]\frac{d\bar{u}_{m1}}{d\zeta}=2\bar{u}_{m1}^2(1-\bar{u}_{m1})^2\left[B_{m1}\left(\frac{d^2\bar{u}_{1c}}{d\eta^2}\right)_0+i_0\kappa_{21}\left(\frac{d^2\bar{u}_{2c}}{d\eta^2}\right)_0(1-B_{m1})\right]-\frac{1}{2}\frac{d\bar{p}}{d\zeta}.$$

The boundary conditions are:

$$\zeta=0,\ B_{m1}=1,\ B_{w1}=B_{w1i},\ \bar{u}_{w1}=\bar{u}_{w1i},\ \bar{u}_{m1}=\bar{u}_{c1i},\ \bar{p}=\bar{p}_i. \tag{32}$$

where $\bar{B}_{1i}=\bar{B}_1(x_i)$, $\bar{B}_{w1i}=\bar{B}_{w1}(x_i)$, $\bar{p}_i=\bar{p}(x_i)$, $\zeta=\kappa_1(x-x_i)/r_0$, $\bar{u}_{m1}=u_{m1}/u_{01}$, $\bar{u}_{c1i}=\sqrt{1-\bar{p}_i}$.

The derived systems of the mixed type equations with corresponding boundary conditions, (29) - (32), represent the mathematical model for two-phase turbulent jet of immiscible liquids in a confined chamber on its total range of flow.

In assumption of a slip of flow on the wall of chamber, the profiles of the main parameters in the form (26) are adopted with the following assignments:

$$u_{mi}=\bar{u}_{mi}u_{mic},\ u_{wi}=\bar{u}_{wi}u_{mic},\ B_{m1}=B_{m1c}\bar{B}_{m1},\ B_{w1}=B_{m1c}\bar{B}_{w1},\ B_{1c}=B_{m1c}\bar{B}_{1c},$$

$p_0$- pressure in a free jet. Then assuming as for free jet $u_{m2}=u_{m1}s_0$, $u_{w2}=u_{w1}s_0$, , we can write the next dimensionless equation array for the turbulent two-phase jet confined at its ground part:

$$\bar{B}_{m1}\left[(\bar{u}_{m1}-\bar{u}_{w1})\int_0^1\bar{B}_{1c}\bar{u}_{1c}\eta d\eta+\bar{u}_{w1}\int_0^1\bar{B}_{1c}\eta d\eta\right]+\bar{B}_{w1}\left[(\bar{u}_{m1}-\bar{u}_{w1})\left(\int_0^1\bar{u}_{1c}\eta d\eta-\int_0^1\bar{B}_{1c}\bar{u}_{1c}\eta d\eta\right)+\right.$$

$$\left.+\left(\frac{1}{2}-\int_0^1\bar{B}_{1c}\eta d\eta\right)\bar{u}_{w1}\right]=\int_0^1\bar{B}_{1c}\bar{u}_{1c}\eta d\eta,\ \bar{B}_{m1}\left[(\bar{u}_{m1}-\bar{u}_{w1})\int_0^1\bar{B}_{1c}\bar{u}_{2c}\eta d\eta+\bar{u}_{w1}\int_0^1\bar{B}_{1c}\eta d\eta\right]+$$

$$+\bar{B}_{w1}\left[(\bar{u}_{m1}-\bar{u}_{w1})\left(\int_0^1\bar{u}_{2c}\eta d\eta-\int_0^1\bar{B}_{1c}\bar{u}_{2c}\eta d\eta\right)+\left(0.5-\int_0^1\bar{B}_{1c}\eta d\eta\right)\bar{u}_{w1}\right]=$$

$$=\int_0^1\bar{B}_{1c}\bar{u}_{2c}\eta d\eta+1/B_{m1c}\left[(\bar{u}_{m1}-\bar{u}_{w1}-1)\int_0^1\bar{u}_{2c}\eta d\eta+0.5\bar{u}_{w1}\right], \tag{33}$$

$$\bar{p}=4B_{m1c}\left\{\int_0^1\bar{B}_{1c}\bar{u}_{1c}^2\eta d\eta+i_0\int_0^1\left(\frac{1}{B_{m1c}}-\bar{B}_{1c}\right)\bar{u}_{2c}\eta d\eta+(B_{m1}-B_{w1})\left[(\bar{u}_{m1}-\bar{u}_{w1})^2\int_0^1\bar{B}_{1c}(i_0\bar{u}_{2c}^2-\bar{u}_{1c}^2)\eta d\eta+\right.\right.$$

$$+2\bar{u}_{w1}(\bar{u}_{m1}-\bar{u}_{w1})\int_0^1 \bar{B}_{1c}(i_0\bar{u}_{2c}^2-\bar{u}_{1c}^2)\eta d\eta+(i_0-1)\bar{u}_{w1}^2\int_0^1 \bar{B}_{1c}\eta d\eta\bigg]+\bar{B}_{w1}\bigg[(\bar{u}_{m1}-\bar{u}_{w1})^2\bigg(i_0\int_0^1 \bar{u}_{2c}\eta d\eta-\int_0^1 \bar{u}_{1c}^2\eta d\eta\bigg)+$$

$$+2\bar{u}_{w1}(\bar{u}_{m1}-\bar{u}_{w1})\bigg(i_0\int_0^1 \bar{u}_{2c}\eta d\eta-\int_0^1 \bar{u}_{1c}\eta d\eta\bigg)+0.5(i_0-1)\bar{u}_{w1}^2\bigg]+$$

$$-i_0/B_{m1c}\bigg[(\bar{u}_{m1}-\bar{u}_{w1})^2\int_0^1 \bar{u}_{2c}^2\eta d\eta+2\bar{u}_{w1}(\bar{u}_{m1}-\bar{u}_{w1})\int_0^1 \bar{u}_{2c}\eta d\eta+0.5\bar{u}_{w1}^2\bigg]\bigg\},$$

$$\frac{d}{d\bar{x}}\bigg\{\int_0^{\eta^*}\big[(\bar{B}_{m1}-\bar{B}_{w1})\bar{B}_{1c}+\bar{B}_{w1}\big]\big[(\bar{u}_{m1}-\bar{u}_{w1})\bar{u}_{1c}+\bar{u}_{w1}\big]^2\eta d\eta+i_0\int_0^{\eta^*}\bigg[\frac{1}{B_{m1c}}-(\bar{B}_{m1}-\bar{B}_{w1})\bar{B}_{1c}-\bar{B}_{w1}\bigg][\bar{u}_{w1}+$$

$$+(\bar{u}_{m1}-\bar{u}_{w1})\bar{u}_{2c}\big]^2\eta d\eta\bigg\}-\big[(\bar{u}_{m1}-\bar{u}_{w1})\bar{u}_{1c}^*+\bar{u}_{w1}\big]\frac{d}{d\bar{x}}\int_0^{\eta^*}\big[(\bar{B}_{m1}-\bar{B}_{w1})\bar{B}_{1c}+\bar{B}_{w1}\big]\big[(\bar{u}_{m1}-\bar{u}_{w1})\bar{u}_{1c}+\bar{u}_{w1}\big]\eta d\eta+$$

$$-i_0\big[(\bar{u}_{m1}-\bar{u}_{w1})\bar{u}_{2c}^*+\bar{u}_{w1}\big]\frac{d}{d\bar{x}}\int_0^{\eta^*}\bigg[\frac{1}{B_{m1c}}+(\bar{B}_{w1}-\bar{B}_{m1})\bar{B}_{1c}+\bar{B}_{w1}\bigg]\big[(\bar{u}_{m1}-\bar{u}_{w1})\bar{u}_{2c}+\bar{u}_{w1}\big]\eta d\eta=$$

$$=\kappa_1\eta^*(\bar{u}_{m1}-\bar{u}_{w1})^2\bigg\{\big[(\bar{B}_{m1}-\bar{B}_{w1})\bar{B}_{1c}^*+\bar{B}_{w1}\big]\bigg[\bigg(\frac{d\bar{u}_{1c}}{d\eta}\bigg)^*-i_0\kappa_{21}\bigg(\frac{d\bar{u}_{2c}}{d\eta}\bigg)^*\bigg]+i_0\kappa_{21}\bigg(\frac{d\bar{u}_{2c}}{d\eta}\bigg)^*\bigg\}-\frac{(\eta^*)^2}{4B_{m1c}}\frac{d\bar{p}}{d\bar{x}},$$

$$\big[\bar{B}_{m1}+i_0(1-\bar{B}_{m1})\big]\bar{u}_{m1}\frac{d\bar{u}_{m1}}{d\bar{x}}=-\frac{1}{2B_{m1c}}\frac{d\bar{p}}{d\bar{x}}+2\kappa_1(\bar{u}_{m1}-\bar{u}_{w1})^2\bigg[\bar{B}_{m1}\bigg(\frac{d^2\bar{u}_{1c}}{d\eta^2}\bigg)_0+i_0\kappa_{21}(1-\bar{B}_{m1})\bigg(\frac{d^2\bar{u}_{2c}}{d\eta^2}\bigg)_0\bigg].$$

The boundary conditions are:

$$\bar{x}=0,\ \bar{u}_{w1}=0,\ \bar{B}_{w1}=0,\ \bar{u}_{m1}=1,\ \bar{B}_{m1}=1. \tag{34}$$

## 8. Computer simulation of the confined jet of two immiscible liquid

Numerical solution of the boundary-value problem (33), (34) has been done for a wide region of the variating parameters. From the algebraic subsystem of the equation array (33) the parameters $\bar{B}_{m1}$, $\bar{B}_{w1}$, $\bar{p}$ are expressed as functions of the velocity components at the jet's axis and at the chamber's wall: $\bar{u}_{m1}$, $\bar{u}_{w1}$. Then the system of two ordinary differential equations for the functions $\bar{u}_{m1}(\bar{x})$, $\bar{u}_{w1}(\bar{x})$ is solved numerically on computer. Some results are presented in Fig. 5:

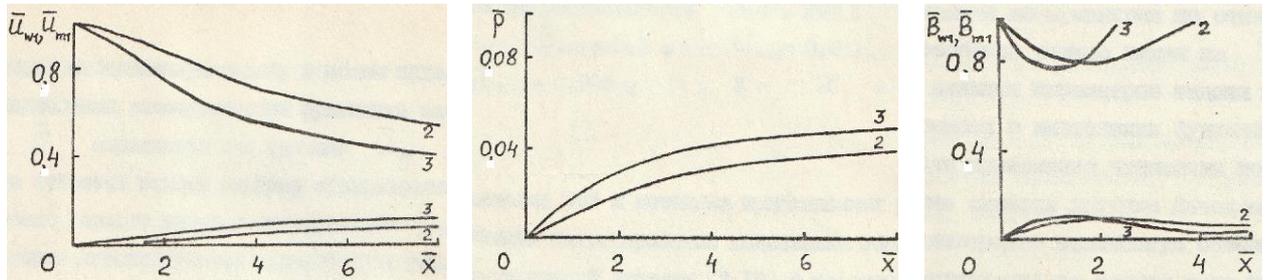

Fig. 5. Velocities, pressure and the functions-indicators for the axis and wall of the chamber:

Here $i_0$=0.3, $B_{m1c}$=0.6, $u_{m1c}$=0.8, $\kappa_{21}$=3.6, two variants: 2- $\kappa_1$=0.006, 3- $\kappa_1$=0.008.

Velocity of a jet at the axis of chamber is slightly falling down (approximately 0.4-0.6) and growing at the wall of a chamber (up to 0.08-0.12). Pressure in a chamber is growing intensively at the entrance and slowly afterwards, nearly linear. The most interesting is peculiarity of the functions-indicators at the axis and at the wall. As clearly seen from the Fig. 5, turbulent mixing in a first phase determines a length of the function variation with the longitudinal coordinate but not the function's character. The higher is turbulent mixing coefficient, the shorter is distance of the varying process but the character and even the value is the same. The first phase at the axis is decreasing shortly at the entrance of the chamber approximately 20%, but then it recovers to the previous level (100% first phase at the axis). Correspondingly, the first phase at the wall of the chamber is growing on the same distance from 0 up to 10% and then falls down below 5%. This calculation was done for the case of lower density of a second phase. Thus, the heavier first phase at the beginning of the chamber is a little distributed across it but then it is nearly totally collected around the axis.

Numerical simulation of the turbulent two-phase jet in a chamber revealed peculiarities different from the above-considered free jet flow. The functions $\bar{u}_{m1}$, $\bar{u}_{w1}$ have strong influence on the $\bar{B}_{m1}$, $\bar{B}_{w1}$, especially strong as concern to $\bar{B}_{w1}$. This cause fast growing of the inaccuracy of calculations because the values $\bar{u}_{m1}$, $\bar{u}_{w1}$ by small errors in numerical solution lead to increased errors in $\bar{B}_{m1}$, $\bar{B}_{w1}$, which are strongly inside the interval from 0 to 1. Such peculiarity of the interconnection of computed parameters cause serious impediments in computer simulations. It is interesting to consider the equilibrium flow in a chamber. As far as the inverse influence of the $\bar{B}_{m1}$, $\bar{B}_{w1}$ to the values $\bar{u}_{m1}$, $\bar{u}_{w1}$ is small, we can write:

$$\bar{x} \to \infty, \ \bar{B}_{m1} \to \bar{B}_{w1} \to \bar{B}_{1\infty}, \ \bar{u}_{m1} \to \bar{u}_{w1} \to \bar{u}_{1\infty}. \tag{35}$$

The condition (35) is about total uniform distribution of the phases and their parameters in a cross-section of a chamber, which may be not achievable. But we can study this question. The following piecewise-linear approximations satisfy to these conditions:

$$\bar{B}_{m1} = 1 + \left(\bar{B}_{1\infty} - 1\right) \sum_{k=1}^{m_1} b_{1k} \bar{x}^k \left(1 + \sum_{l=1}^{m_1} b_{2l} \bar{x}^l\right)^{-1}, \quad \bar{B}_{w1} = \bar{B}_{1\infty} \sum_{j=1}^{m_2} t_{1j} \bar{x}^j \left(1 + \sum_{l=1}^{m_2} t_{2i} \bar{x}^i\right)^{-1}, \tag{36}$$

where $b_{1k}$, $b_{2l}$, $t_{1j}$, $t_{2i}$ are constants computed from the equations (33). In general, their number is $2(m_1 + m_2)$.

As far as the above task of finding the coefficients is very cumbersome already in a first approach, it was proposed accounting a weak influence of the $\bar{B}_{m1}$, $\bar{B}_{w1}$ on $\bar{u}_{m1}$, $\bar{u}_{w1}$, adopt for the functions $\bar{B}_{m1}$, $\bar{B}_{w1}$ the following simple approximations:

$$\bar{B}_{m1} = 1 + \left(\bar{B}_{1\infty} - 1\right) \frac{b_1 \tilde{x}}{1 + b_2 \tilde{x}}, \qquad \bar{B}_{w1} = \bar{B}_{1\infty} \frac{t_1 \tilde{x}}{1 + t_2 \tilde{x}}. \tag{37}$$

We state the following conditions:

$$\tilde{x} = 1, \ \bar{B}_{m1} = \bar{B}_{w1} = \bar{B}_{1\infty}; \quad \tilde{x} = 0.5, \ \bar{B}_{m1} = 1.01 \bar{B}_{1\infty}, \ \bar{B}_{w1} = 0.99 \bar{B}_{1\infty}, \tag{38}$$

where $\tilde{x} = \chi \bar{x}$, $\chi = \chi(i_0, x_c, \bar{x})$ - some empirical function, e.g. $\chi = \chi(i_0, x_c)$ = const, in particular.

Then the following assignments are introduced:

$$\text{AI5} = \int_0^1 \bar{B}_{1c} \bar{u}_{1c} \eta d\eta, \ \text{AI8} = \int_0^1 \bar{u}_{1c} \eta d\eta, \ \text{AI11} = \int_0^1 \bar{u}_{1c}^2 \eta d\eta, \ \text{AI14} = \int_0^1 \bar{B}_{1c} \bar{u}_{1c}^2 \eta d\eta, \tag{39}$$

$$\text{AI15} = \int_0^1 \bar{B}_{1c} \bar{u}_{2c}^2 \eta d\eta, \ \text{AI16} = \int_0^1 \bar{B}_{1c} \bar{u}_{2c} \eta d\eta, \ \text{AEP} = \int_0^1 \bar{u}_{2c}^2 \eta d\eta, \ \text{BT} = \int_0^1 \bar{u}_{2c} \eta d\eta, \ \text{PL} = \int_0^1 \bar{B}_{1c} \eta d\eta.$$

Now from the first 3 equations of the system (33), the following expressions yield for the above-introduced parameters:

$$\bar{u}_{1\infty} = 2[BT + B_{m1c}(AI5-AI16)], \quad \bar{B}_{1\infty} = 2AI5/\bar{u}_{1\infty}, \quad \bar{u}_{m1} = (1 - F_2 \bar{u}_{w1}), \quad \bar{u}_{w1} = (F_1 F_5 - F_3)/(F_1 F_4 - F_2 F_3),$$

$$\bar{p} = 4B_{m1c}\{AI14 + i_0(AEP/B_{m1c} - AI15) + (\bar{B}_{m1} - \bar{B}_{w1})[(\bar{u}_{m1} - \bar{u}_{w1})^2(i_0 AI15 - AI14) +$$
$$+ 2(\bar{u}_{m1} - \bar{u}_{w1})\bar{u}_{w1}(i_0 AI16 - AI15) + (i_0 - 1)PL\bar{u}_{w1}^2] + \bar{B}_{w1}[(\bar{u}_{m1} - \bar{u}_{w1})^2(i_0 AEP - AI11) + \qquad (40)$$
$$+ 2\bar{u}_{w1}(\bar{u}_{m1} - \bar{u}_{w1})(i_0 BT - AI8) + 0.5(i_0 - 1)\bar{u}_{w1}^2] - \frac{i_0}{B_{m1c}}\left[\frac{\bar{u}_{w1}^2}{2} + (\bar{u}_{m1} - \bar{u}_{w1})^2 AEP + 2\bar{u}_{w1}(\bar{u}_{m1} - \bar{u}_{w1})BT\right]\},$$

where are: $F_5 = AI16 - BT/B_{m1c}$, $F_4 = \bar{B}_{m1}(PL - AI16) + \bar{B}_{w1}(0.5 - PL - BT + AI16) + (BT - 0.5)/B_{m1c}$, $F_3 = \bar{B}_{m1} AI16 + \bar{B}_{w1}(BT - AI16) - BT B_{m1c}$, $F_2 = \bar{B}_{m1}(PL/AI5 - 1) + \bar{B}_{w1}[1 + (0.5 - PL - AI8)/AI5]$, $F_1 = \bar{B}_{m1} + \bar{B}_{w1}(AI8/AI5 - 1)$.

The obtained numerical solution of the boundary problem (33), (34) depends only on one empirical constant or function $\chi$, which is multiplayer in the variable $\tilde{x} = \chi \bar{x}$ and does not depend on the constants of turbulent mixing in the phases $\kappa_i$. Therefore, it does not require an introduction of any hypotheses concerning the turbulent shear stress. The correlation of $\chi$ with empirical constants of the turbulent mixing $\kappa_i$ is easily computed from the solution of the two last equations of the system (33).

### 9. Approximate numerical solution for the totally stabilized confined jet flow

The parameters of the totally stabilized flow of the confined two-phase turbulent jet of two immiscible liquids (e.g. $\bar{B}_{1\infty}$ and the others) computed from the first three equations (33), and the constants $t_i$, $b_i$ determined from substitution of (37) into (38), are given in the Table 1:

Table 1. Parameters of the stabilized confined jet

| $i_0$ | 0,3 | | | 1,0 | | | 8,0 | | |
|---|---|---|---|---|---|---|---|---|---|
| $B_{m1c}$ | 1,0 | 0,8 | 0,6 | 1,0 | 0,8 | 0,6 | 1,0 | 0,8 | 0,6 |
| $u_{m1c}$ | 1,0 | 0,92 | 0,8 | 1,0 | 0,84 | 0,66 | 1,0 | 0,53 | 0,29 |
| $R$ | 3,48 | 4,11 | 5,0 | 2,75 | 3,35 | 4,38 | 2,09 | 3,18 | 5,0 |
| $\bar{u}_{1\infty}$ | | 0,21 | | | 0,22 | | 0,27 | 0,25 | 0,24 |
| $\bar{B}_{1\infty}$ | 0,39 | 0,39 | 0,4 | 0,58 | 0,60 | 0,61 | 0,87 | 0,91 | 1,0 |
| $\bar{p}_\infty$ | 0,11 | 0,12 | 0,14 | 0,16 | 0,16 | 0,15 | 0,19 | 0,34 | 0,51 |
| $b_1$ | 156 | 153 | 150 | 70 | 60 | 62 | 14 | 8,5 | 0 |
| $b_2$ | 155 | 152 | 149 | 69 | 59 | 61 | 13 | 7,5 | 0 |
| $t_1$ | | 99 | | | 99 | | | 99 | |
| $t_2$ | | 98 | | | 98 | | | 98 | |

The three cases $i_0 = 0.3$, $B_{m1c} = 0.6$, $u_{m1c} = 0.8$, were analyzed as previously. Dimensionless velocity of the totally stabilized flow independently of $i_0$ and other parameters is close to 0.2, while the function-indicator of the phase is substantially depending on the parameter $i_0$ (density ratio multiplied by square of the slip ratio for the phases). Dimensionless pressure depends a little on a density ratio, which abruptly increases only by big density ratio and a small velocity of the jet's entrainment into a chamber. The last case may evidence a possibility for jamming a jet in a chamber. The obtained approximate solution for a spreading of the turbulent two-phase jet in a chamber is presented in Fig. 6, where the velocities of a flow and pressure are shown against axis of the chamber. It is clearly observed that the most intensive mixing in a flow is going by $\tilde{x} \leq 0.1$, afterward it is going slower and slower, especially concerning the pressure. Presence of a maximum of recoverable pressure by high density ratio allows deciding in a frame of the model, in a first approach, about the rational length of the stabilizing chamber.

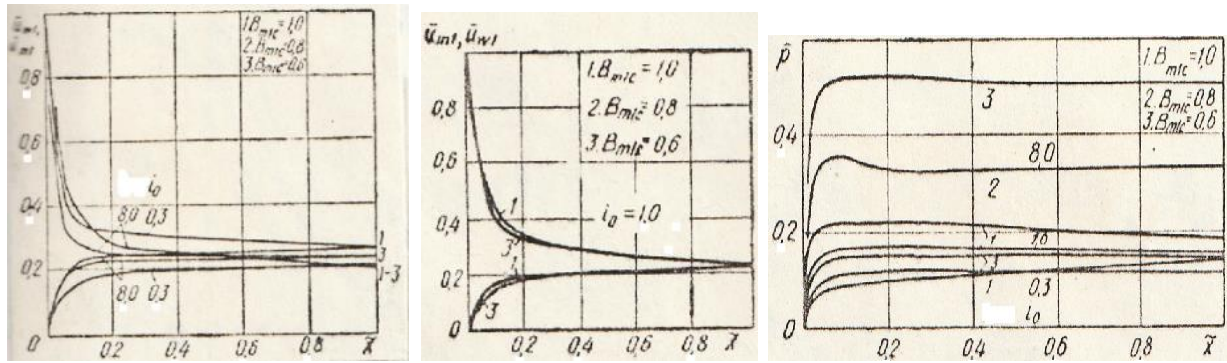

Fig. 6. Velocity and pressure of a jet flow in a chamber against axial coordinate

Provement of the results obtained by numerical solution of the boundary problem (33), (34) and its approximate solution above with the known experimental data was done. By two-phase flows such data are absent; therefore a comparison with the experiments by a turbulent mixing in the jet devices working on homogeneous flows of incompressible liquids was done (Fig. 7).

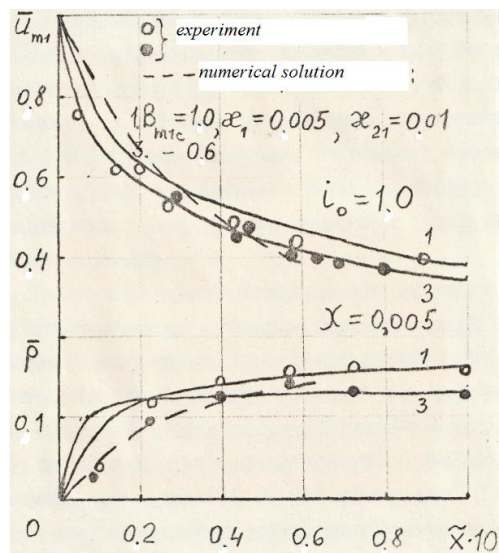

Fig. 7. Comparison of the solutions obtained with experimental data

As seen from Fig. 7, for $\chi = 0.005$, the correspondence of the results obtained with experimental data [19] is good. It may be improved with choosing the function $\chi(\overline{x})$ for the best correlation in each specific case.

## 10. Application of the method for modeling of corium cooling during severe accidents at NPP

We applied the above-described mathematical model for numerical simulation of the two-phase turbulent flow of melt with vapor for a cooling of the corium melt during severe accidents at the nuclear power plants (NPP) [13] as shown in Fig. 8:

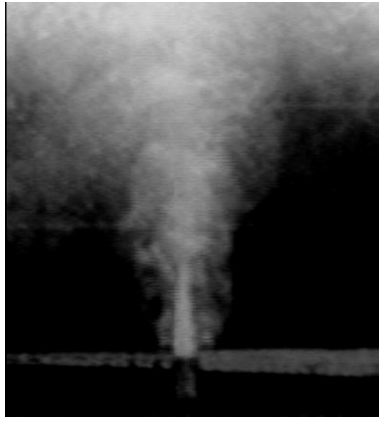

Fig. 8. Jet of volatile coolant penetrating the pool of melt from below

As was shown above, for $i_0>3$, $h$ is in the range from -6 to 0, therefore the approximation (4) was used for the function-indicator of the first phase in all range of parameters for the considered task. The constants for the integrals in our integral correlations were as follows

$$\begin{pmatrix} a_{11} & a_{12} & b_{11} & b_{12} \\ a_{21} & a_{22} & b_{21} & b_{22} \\ a_{31} & a_{32} & b_{31} & b_{32} \\ a_{41} & a_{42} & b_{41} & b_{42} \end{pmatrix} = \begin{pmatrix} 0.5464 & 0.0208 & 0.0179 & -0.0101 \\ 0.1667 & 0.0101 & 0.0095 & -0.0042 \\ 0.4604 & 0.0139 & 0.0054 & -0.0048 \\ 0.1198 & 0.0059 & 0.0023 & -0.0016 \end{pmatrix}$$

By these parameters accounting that $i_0 \gg 3$ it was calculated from (8) for the initial part of a jet:

$$h_0 = h(0) = \frac{0.0054 i_0 - 0.0860}{0.0069 + 0.0048 i_0}, \qquad h_i = h(\varsigma_i) = \frac{0.0023 i_0 - 0.0469}{0.0016 i_0 + 0.0042}. \tag{41}$$

we estimated that because the maximal value for $h$ is 0 ($B_1$ cannot exceed 1), it yields: $h_0=0$, max $i_0=16$ approximately. And by this value of $i_0$ the slip of phases is estimated as $s_0 = 4\sqrt{\rho_{12}}$, so that if $\rho_{21}=400$, then $u_{02}=2u_{01}$ for the initial part of the jet. And in a similar way for the end of the initial part of the jet: $h_1=0$, max $i_0=20.4$ and $s_0 = 4.52\sqrt{\rho_{12}}$, so that the difference in the parameter $s_0$ between the begin and end of the initial part of the jet is about 13%.

The volumetric ejection coefficient and the expansion rate for the end of the initial part of a jet are the following: $q_1=0.013$, $\delta_1=1.73$. As it is shown the entrainment is very small and the expansion is less than 2. For the begin of ground part of the jet from (11) yields:

$$\begin{pmatrix} \alpha_{11} & \alpha_{12} & \beta_{11} & \beta_{12} \\ \alpha_{21} & \alpha_{22} & \beta_{21} & \beta_{22} \end{pmatrix} = \begin{pmatrix} 0.1286 & 0.0071 & -0.0905 & -0.0040 \\ 0.0792 & 0.0033 & -0.0501 & -0.0016 \end{pmatrix}$$

$\beta_{10}=0.1$, $\beta_{20}=0.052$, and then

$$h_t = \frac{0.0023i - 0.0494}{0.0016i + 0.0038}, \qquad \delta_t = \frac{1}{\sqrt{2(0.0071 h_t + 0.1286)}}, \tag{42}$$

The maximal $h_t=0$, therefore max $i=21.48$ and max $s = 4.64\sqrt{\rho_{12}}$, or by the above-mentioned parameters max $s=0.232$, $\delta_t=1.97$. And from (10), (12) yields the following boundary problem for the ground part of the jet:

$$B_{m1} = \frac{1.13 \bar{u}_{m1}}{0.13 + \bar{u}_{m1}}, \qquad \delta = \frac{1.97}{\bar{u}_{m1}} \sqrt{\frac{0.13 + \bar{u}_{m1}}{1.13}}, \qquad \frac{d\bar{u}_{m1}}{d\zeta} = -\frac{6.46 \bar{u}_{m1}^2}{\sqrt{0.13 + \bar{u}_{m1}}} \frac{\bar{u}_{m1} + 5\kappa_{21}(1-\bar{u}_{m1})}{2.5 - 1.5\bar{u}_{m1}}. \tag{43}$$

This boundary problem was solved numerically for $\kappa_{21} \in [0.1, 1.0]$.

Simulation of the cooling process of a melt pool with volatile coolant the integral entrainment of the melt into a mixing layer of coolant $Q = 1/H \int_0^H q(x)dx$, where $H$ is the height of the pool. In a numerical solution:

$$Q = \frac{1}{H} \sum_{i=1}^{N} q(x_i)(x_i - x_{i-1}), \qquad (44)$$

where $q(x_i)$ is the ejection coefficient, $x_i$ is current point in the partition of the calculated interval by $x$. The time for vapor through the pool is $\Delta t = H/u$, $u$ is the mean velocity of vapor in a mixing layer:

$$u = (1/H) \sum_{i=1}^{N} u_{mi}(x_i - x_{i-1}). \qquad (45)$$

Assuming that every portion of coolant vaporizes near the nozzle, one can get a heat removal from a melt by a coolant according to the following heat balance equation:

$$c_2 M_2 (T_{20} - T_2) = q_c \Delta t h_{lv}, \qquad (46)$$

where $q_c$ is the coolant's flow rate, $h_{lv}$ is a heat of vaporization, $T_{20}$, $T_2$ are initial and current temperature of a melt, respectively, $M_2$ is a mass of the melt in a pool to be cooled down by volatile coolant supplied by the nozzles. From the above equation yields the following reduced melt's temperature:

$$T_2 = T_{20} - q_c \Delta t h_{lv} / (c_2 M_2). \qquad (47)$$

Then the heat balance for one cycle of a vapor portion going from the nozzle to a top of the pool is expressed as $c_2 \rho_2 QV(T_2 - T) = c_1 \rho_1 (1-Q) V (T - T_1)$, where $V$ is the total volume of a mixing zone and $T$ is the equilibrium temperature after heat exchange. From the last correlation it is obtained

$$T = \frac{c_2 \rho_2 Q T_2 + c_1 \rho_1 (1-Q) T_1}{c_2 \rho_2 Q + c_1 \rho_1 (1-Q)}. \qquad (48)$$

Assuming that a vapor is going off the pool from a free surface of it and the drops are falling down into a pool, we can continue with account also the melt's cooling with such drops: $m_t T + (M_2 - m_t) T_2 = M_2 T_2$, where $m_t$ is the mass of the drops in a jet, which is as follows:

$$m_t = 0.25 \pi Q \rho_2 (\bar{\delta} + 1)^2 H r_0^2. \qquad (49)$$

Afterward, from (48), the correlated temperature of a melt due to one circle of a jet passage through the melt pool is got. Step-by-step, the temperature of a melt with time was calculated. According to the above, $n=0.27$, $\kappa_{21}=0.6$; $n=1.0$, $\kappa_{21}=0.4$; $n=7.3$, $\kappa_{21}=0.2$ were computed [13]. The next approximation was done:

$$\kappa_{21} = 0.4 - 0.1326 \ln n + 0.0163 (\ln n)^2, \qquad (50)$$

where from the minimal value of $\kappa_{21}$ was 0.13. For big values of $n$ in this case it was stated $\kappa_{21}=0.1$.

### 11. Validation of the method for corium cooling during severe accidents at NPP

The above model was used to calculate the characteristics of the process of cooling the melt's pool with the jets of volatile coolant supplied from the bottom. The jets of liquid coolant due to high temperature of the melt in a pool are vaporizing close to the exit from the nozzle so that further the vapor jets are mixing with the melt in a pool cooling it down. The thermal parameters governing the boiling process have been studied in [20]. First, the averaged values of $\kappa_1$ were obtained from the experiments (see Table 2), then by the above model the characteristic parameters were computed. As seen from the results [13] presented in Figs 9, 10, the method is good for such complex two-phase jets, which allows computing important characteristics.

Table 2. Experimental data: $r_0$=0.3 mm, $\mu_1 = 2 \cdot 10^{-5} N \cdot s / m^2$, $\rho_1 = 2.1 kg / m^3$

| Cases | $\kappa_1$ | | $u_{0,l}$ (m/sec) | | $\rho_{2,1}$ | | $\mu_{2,1}$ (20°C) | |
| --- | --- | --- | --- | --- | --- | --- | --- | --- |
| | water | paraf. | water | paraf. | water | paraf. | water | paraf. |
| (a) | 0.008 | 0.016 | 0.23 | 0.29 | 476 | 419 | 4.16 | 271 |
| (b) | 0.031 | 0.007 | 0.29 | 0.59 | 476 | 419 | 4.16 | 271 |
| (c) | 0.058 | 0.014 | 0.59 | 0.17 | 476 | 419 | 4.16 | 271 |
| (d) | 0.014 | 0.009 | 0.29 | 0.29 | 476 | 419 | 4.16 | 271 |
| (e) | 0.008 | 0.014 | 0.59 | 0.47 | 476 | 419 | 4.16 | 271 |
| (f) | 0.012 | 0.011 | 0.29 | 0.59 | 476 | 419 | 4.16 | 271 |
| (g) | 0.012 | 0.011 | 0.59 | 0.77 | 476 | 419 | 4.16 | 271 |
| (h) | - | 0.012 | - | 1.01 | - | 419 | - | 271 |

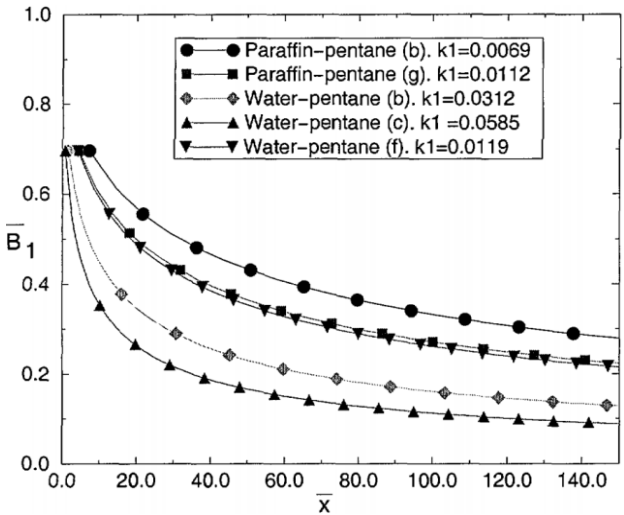
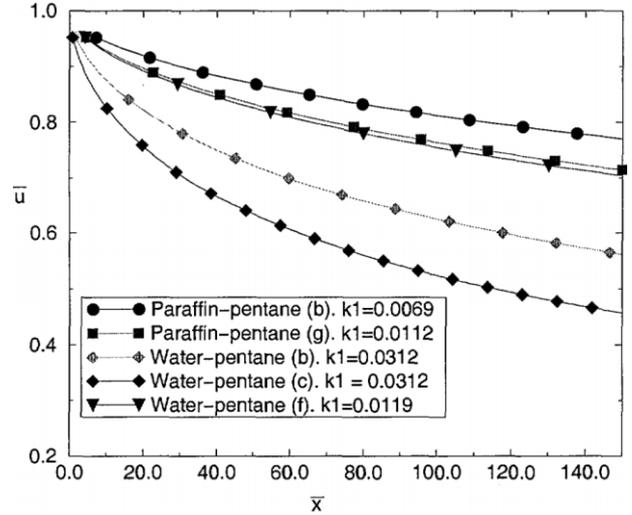

Fig. 9 Tendency for the function-indicator $B_1$ and flow velocity for different flow cases

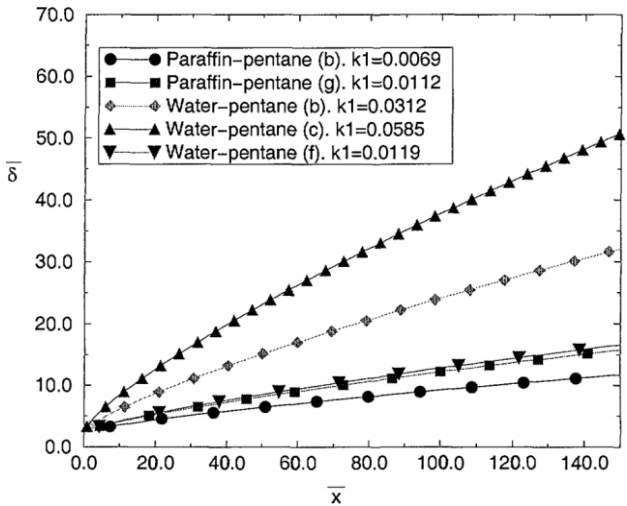
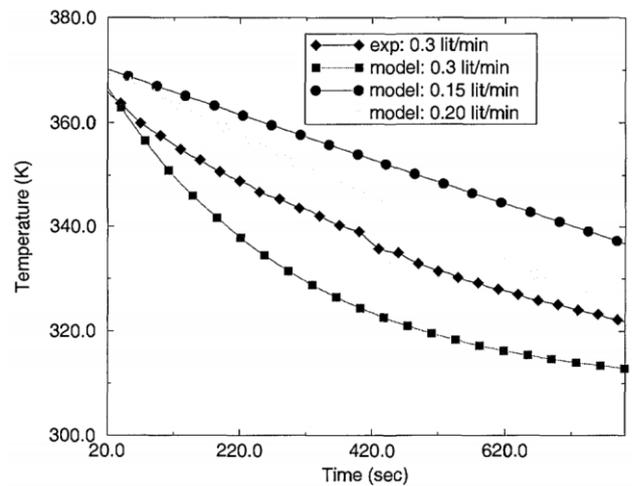

Fig. 10. The width of the mixing zone and the temperature profile for the model and experiment

The conducted study of the mixing processes in turbulent heterogeneous jet of immiscible liquids in jet devices allows determining the main parameters of the devices including distribution of the parameters along the axis of the devices, as well as across the mixing layer. This is important for optimal organization of the working process. The results may be useful in a number of chemical technology and other engineering fields, where jet hydraulic machines are applied. The mathematical model developed for the free and confined jets of two-phase flows and the approximate correlations proposed from analysis are available for implementation into the research and engineering calculations by other scientists and engineers.

The problem of turbulent mixing in jet machines working on mutually immiscible liquids is quite complex and not so good investigated. Therefore, this attempt can be interesting both, in theoretical and practical aspects because it revealed some basic features of the multiphase flow. Later on, the processes will be studied more in deep. The data obtained are expected to improve the performance and quality of the jet machines in different industrial and technical applications. The distribution of the phases in a mixing flows are highly important for deep understanding and optimization of mixing processes.